# On the electrochemical $CO_2$ reduction by Bi-based catalysts: single crystals or mixture phases

Mengting Zhou,[a,†] Hongxia Liu,[b,†] Juntao Yan,[c] Qingjun Chen,[d] Rong Chen,[e] Lei Liu[b]

[a] School of Chemistry and Environmental Engineering, Wuhan Institute of Technology, 430205 Wuhan, PR China

[b] Center for Computational Chemistry, School of Chemistry and Chemical Engineering, Wuhan Textile University, Wuhan 430200, China

[c] College of Chemistry and Environmental Engineering, Wuhan Polytechnic University, Wuhan 430023, China

[d] Key Laboratory of Rare Earths, Chinese Academy of Sciences, Ganjiang Innovation Academy, Chinese Academy of Sciences, Ganzhou 341000, China

[e] State Key Laboratory of New Textile Materials & Advanced Processing Technologies, Wuhan Textile University, 430200 Wuhan, China

† These authors contributed equally
* Corresponding authors:
Rong Chen: rchenhku@hotmail.com
Lei Liu: liulei@wtu.edu.cn; liulei3039@gmail.com

**ABSTRACT**

Metallic bismuth is both non-toxic and cost-effective. Bi-based catalysts have demonstrated the ability to efficiently produce HCOOH through $CO_2RR$ while effectively inhibiting the HER. Although many experiments have been reported concerning its performance towards $CO_2$ reduction, the impact its valence states and crystal faces on $CO_2RR$ selectivity (e.g. HCOOH versus CO) it still under debate. Here, we performed a comprehensive study *via* density functional theory, by including three typical valence states of Bi, such as 0 (Bi), +3 ($Bi_2O_3$) and +5 ($Bi_2O_5$), as well as their often-studied crystal facets. The results show that metallic Bi demonstrates a poor selectivity for HCOOH, but boasts a higher conversion rate for $CO_2$. While $Bi_2O_3$ exhibits a good selectivity for HCOOH production, yet it displays a lower conversion rate for $CO_2$. For $Bi_2O_5$, all studied surfaces show high energy barriers in both cases of HCOOH and CO production, and lower energy barriers for HER reactions, indicating that Bi at +5 valence state is not the good choice for 2*e* transfer reactions. Subsequently, we further examined the effects of oxygen contents on the selectivity of HCOOH and the conversion rate for $CO_2$. Interestingly, we found that partial oxidization of Bi benefits both the selectivity and the conversion rate. With these observations, we suggest that a mixture of Bi (0) and $Bi_2O_3$ (+3) phases would be a better choice than single crystals for future experiments.

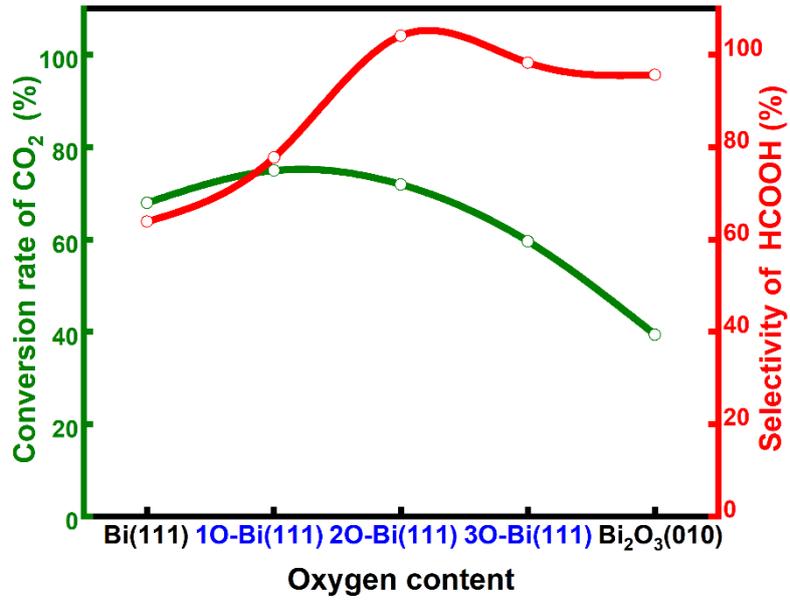

**TOC:** The CO$_2$RR on bismuth-based catalysts were analyzed via DFT calculations, and the results show that the selectivity of HCOOH and the conversion rate of CO$_2$ is strongly depended on the valence states of bismuth.

1. **INTRODUCTION**

With the rapid development of social economy, $CO_2$ emissions continue to increase, which inevitably leads to many environmental issues.[1, 2] Recently, electrochemical $CO_2$ reduction ($CO_2$RR) becomes an important method to achieve carbon neutrality,[3] which can reduce $CO_2$ into various valuable chemical substances. Among them, formation of CO and HCOOH only involves two electron transfer processes, and their conversion efficiency could approach nearly 99% .[4, 5] Moreover, as an important liquid raw material in chemical production, HCOOH is widely used in many fields, such as chemical industry, medicine production, and industrial solvents.[6-8] Often noble metals including Pb, Hg, In, Sn and Cd have been employed to catalyze the production of HCOOH in the process of $CO_2$RR. [9-12] Recently, bismuth (Bi), a non-toxic and inexpensive metal catalyst, has been reported that not only generates HCOOH in the $CO_2$RR process with high yields, but also has a high overpotential against hydrogen evolution reactions (HER).[13-18] Subsequently, Bi oxide and Bi contained catalysts have been widely investigated toward their properties of HCOOH production *via* $CO_2$RR techniques. [19-29]

Normally, it is difficult for experiments to make pure Bi (the +0 valence state of Bi) catalysts, as it is would be easily oxidized to their high valence states. [30] Hence, several indirect methods have been employed, for example, Han et al. reported that an ultra-thin BiOI nanosheet was used as a template, then the halogen layers was removed, and the bismuth oxide was reduced at cathode, generating Bi (001) crystal surface for $CO_2$RR to produce HCOOH. Note that the scanning electron microscope（SEM）image showed that overall 2D structure was remained unchanged before and after the reactions .[31] Arquer et al. used BiOBr as a template, which was reduced to Bi ($1\bar{1}0$) at the cathode, and the current density was significantly higher than that of Bi (001) crystal surface. Interestingly, the selectivity of HCOOH with BiOBr-template catalyst reaches 90%, while the metallic Bi only have a selectivity of HCOOH being 70%.[32] The crystal structure of BiOBr templated catalysts was shown consisting of a combination of Bi and $Bi_2O_3$ species. These observations indicate that valence states of Bi have strong effects on their catalytic performance of $CO_2RR$. Moreover, the valence state could be even inter-changed during the $CO_2RR$ process. For example, based on flow cell, transmission electron microscope (TEM) and operando Raman spectroscopy, Beatriz et al. found the internal recombination of $Bi_2O_3$ nanoparticles into Bi nanosheets during $CO_2RR$.[33] Li et al. also proved that during the $CO_2RR$ process, the surface active site of Bi-based catalysts is mainly the metallic Bi(0), which is the result of re-construction.[34] Moreover, the experiments identified that normally only the surface part of the high valence state Bi is reduced in the process of $CO_2RR$, while the bulk phase part still retains the structure of the hypervalent state.[35]

To understand above-mentioned experimental results, and to gain deep insights into the effects of valence states on the selectivity of $2e^-$ transfer products, here, we performed following density functional theory studies: firstly, we selected three typical Bi-based catalysts, *i.e.* bismuth (+0), $Bi_2O_3$ (+3) and $Bi_2O_5$ (+5), constructed several low-index crystal surfaces as catalytic surfaces, and studied the efficiency of $CO_2RR$ to produce HCOOH and CO; Secondly, we constructed Bi (111) crystal surfaces with different oxidation degrees, and analyzed their performance towards selectivity of HCCOH versus CO. The results show that metallic Bi demonstrates a poor selectivity for HCOOH, but have a higher conversion rate for $CO_2$. While $Bi_2O_3$ exhibits good selectivity for HCOOH production, yet it displays a lower conversion rate for $CO_2$. Importantly, we found that a mixture of 0 and +3 valence states (or partial oxidation of metallic Bi) gives the best balance between the selectivity of HCOOH and the conversion rate of $CO_2$. We believe that such findings would benefits for the further experiments in the selection and synthesis of Bi based catalysts for $CO_2RR$.

## 2. COMPUTATIONAL DETAILS

### 2.1. Computational methods

All density functional theory (DFT) calculations were performed by using the Vienna Ab initio Simulation Package (VASP), version 6.4.2.[36, 37] The PBE functional in the framework of the generalized gradient approximation (GGA) was used[38, 39], together with the Projector Augmented Wave (PAW) pseudopotential framework.[32] The Gamma point was used to ensure the symmetry of the K-point grid in the Brillouin zone. The DFT-D3 correction method proposed by Grimme et al. was used to describe the weak interactions between molecules and surfaces.[40] The plane wave base group (PW) was used to fit the wave function of the electrons in the periodic system, with a cutoff energy of 450 eV. In all the calculations, the total energy and force convergence were set to be $1\times10^{-5}$ eV, and 0.02 eV/Å, respectively. The spin polarization and a Gaussian broadening of 0.2 eV was employed during the calculations. The formulas to compute surface energy ($E_{surface}$) and formation energy ($E_{formation}$) are as follows:

$$E_{surface} = \frac{E_{slab} - NE_{bulk}}{2A} \quad (Eq.\ 1)$$

$$E_{formation} = \frac{E_{slab} - NE_{bulk}}{N} \quad (Eq.\ 2)$$

in which, $E_{bulk}$ represents energies for bulk systems, $E_{slab}$ represents t energies for different crystal faces, $A$ represents the surface area of slabs, and $N$ represents the number of layers.

The elementary reaction steps for HCOOH formation are as follows:

$$* + CO_{2(g)} \rightarrow CO_2^* \quad (Eq.\ 3)$$

$$*CO_2 + H^+_{(aq)} + e^- \rightarrow OCHO^* \quad (Eq.\ 4)$$

$$OCHO^* + H^+_{(aq)} + e^- \rightarrow HCOOH^* \quad (Eq.\ 5)$$

$$HCOOH^* \rightarrow HCOOH + * \quad (Eq.\ 6)$$

The elementary reaction steps CO formation are as follows:

$$* + CO_{2(g)} \rightarrow CO_2^* \quad (Eq.\ 7)$$

$$*CO_2 + H^+_{(aq)} + e^- \rightarrow COOH^* \quad (Eq.\ 8)$$

$$COOH^* + H^+_{(aq)} + e^- \rightarrow CO^* + H_2O \qquad (Eq.\ 9)$$

$$CO^* \rightarrow CO + * \qquad (Eq.\ 10)$$

### 2.2. Computational models

### 2.2.1. Surface structure of Bi (0)

Metallic Bi was used to study the effects of 0 valence state, denoted as Bi (0), on the performance of $CO_2RR$. Three commonly reported low-index crystal faces (100), (110) and (111) of metallic Bi were selected to represent the surface catalytic structures.[19, 41, 42] The lattice constants of the optimized primitive cell of metallic Bi are $a=b=4.66$ Å, and $c=11.38$ Å, which are well consistent with the experimental data, where $a=b=4.54$ Å, and $c=11.83$ Å.[41] To establish the surface structures sufficient for the adsorption of intermediate molecules, three low-index crystal faces were described by slab models including 36, 40 and 60 Bi atoms, respectively. The lattice parameters after structure optimization were $a=9.32$ Å, $b=11.38$ Å, $a=13.16$ Å, $b=9.30$ Å, and $a=16.14$ Å, $b=12.30$ Å, for (100), (110) and (111), respectively (see Figure 1 for the top and side views of the optimized slabs).

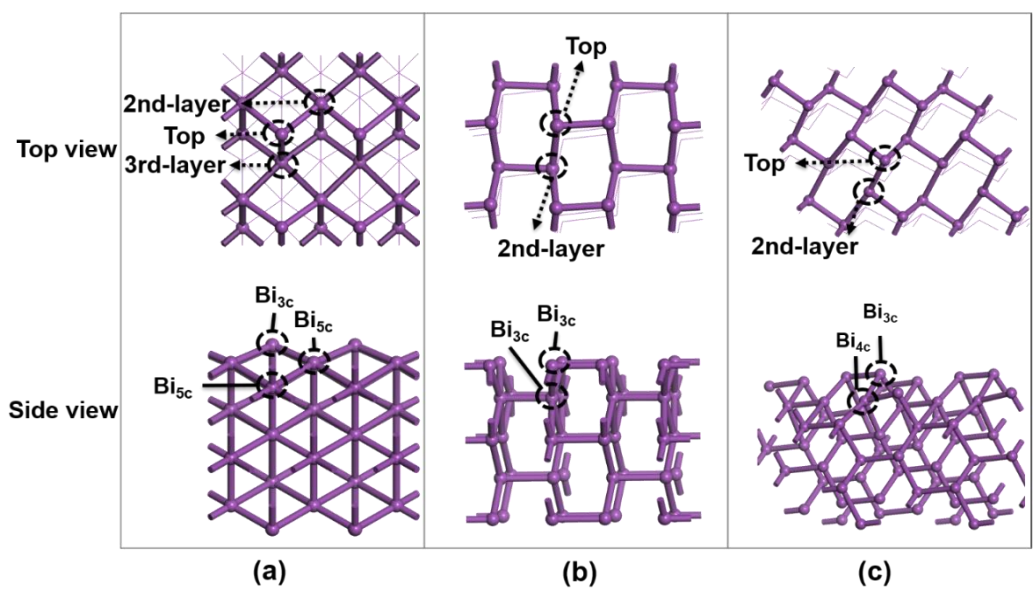

**Figure 1.** Surface morphology and the corresponding adsorption sites of metallic Bi slabs: (a) (100), (b) (110), and (c) (111), respectively. Color legend: Bi purple.

**2.2.2. Surface structure of Bi (+3)**

$Bi_2O_3$ was used to study the effects of +3 valence state, denoted as Bi (+3), on the performance of $CO_2RR$. Three commonly reported low-index crystal faces, such as (100), (010) and (001) of $Bi_2O_3$, were selected to represent the surface catalytic structures.[43] The lattice constants of bulk $Bi_2O_3$ after optimization are $a$= 5.87 Å, $b$=8.21 Å, and $c$= 7.49 Å, which are well consistent with the experimental data ($a$=5.84 Å, $b$= 8.16 Å, and $c$= 7.5 Å).[25, 44] To establish the surface structures sufficient for the adsorption of intermediate molecules, three low-index crystal faces are described by slab models including 48 O atoms and 32 Bi atoms. The lattice parameters after structure optimization are $a$= 16.43 Å, $b$= 14.98 Å, $a$= 14.98 Å, $b$= 11.75 Å, $a$= 11.75 Å, $b$= 16.43 Å, respectively (see Figure 2 for the top and side views of the optimized structures).

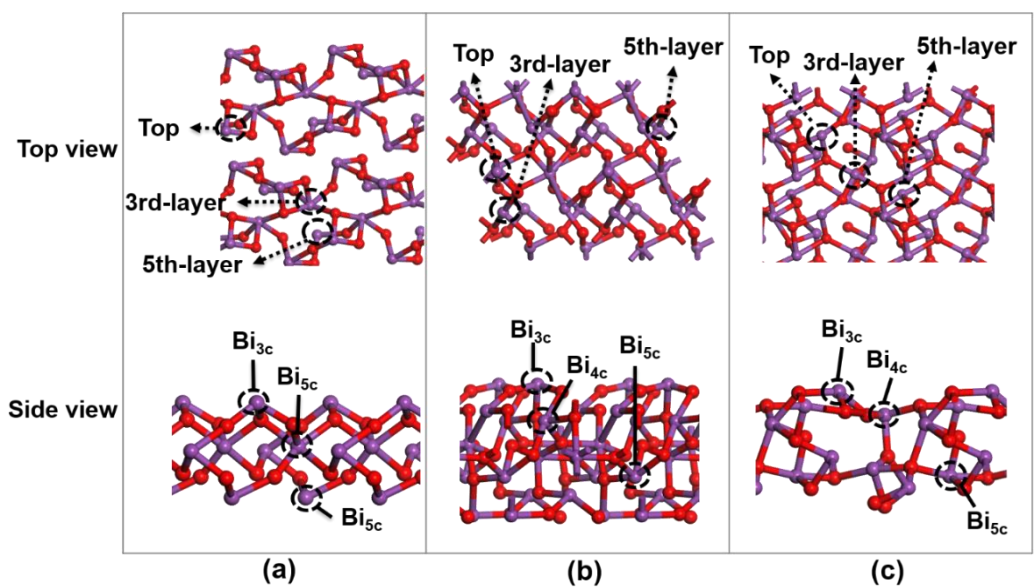

**Figure 2**. Surface morphology and the corresponding adsorption sites of $Bi_2O_3$: (a) (100), (b) (010), and (c) (001) surfaces, respectively. Color legend: Bi purple, and O red.

**2.2.3. Surface structure of Bi (+5)**

$Bi_2O_5$ was used to study the effects of +5 valence state of Bi, denoted as Bi (+5), on the performance of $CO_2RR$. Five commonly reported low-index crystal faces, such as (100), (110), (111), (010) and (001) of $Bi_2O_5$, were selected to represent the surface catalytic structures. The lattice parameters of bulk $Bi_2O_5$ after optimization are $a$= 4.22 Å, $b$= 11.00

Å, and $c$= 10.03 Å, which are well consistent with the experimental data, where $a$= 4.26 Å, $b$= 11.11 Å, and $c$= 10.1921 Å. To establish the surface structure sufficient for the adsorption of intermediate molecules, five low-index crystal faces are described by slab models including 48 O atoms and 32 Bi atoms. The lattice parameters after structure optimization are $a$= 11.00 Å, $b$= 10.03 Å, $a$= 10.03 Å, $b$= 11.79 Å; $a$= 11.79 Å, $b$= 10.88 Å, $a$= 11.79 Å, $b$= 12.69 Å, and $a$= 10.03 Å, $b$= 12.69 Å for (100), (110), (111), (010) and (001), respectively (see Figure 3 for top and side views of the optimized structures).

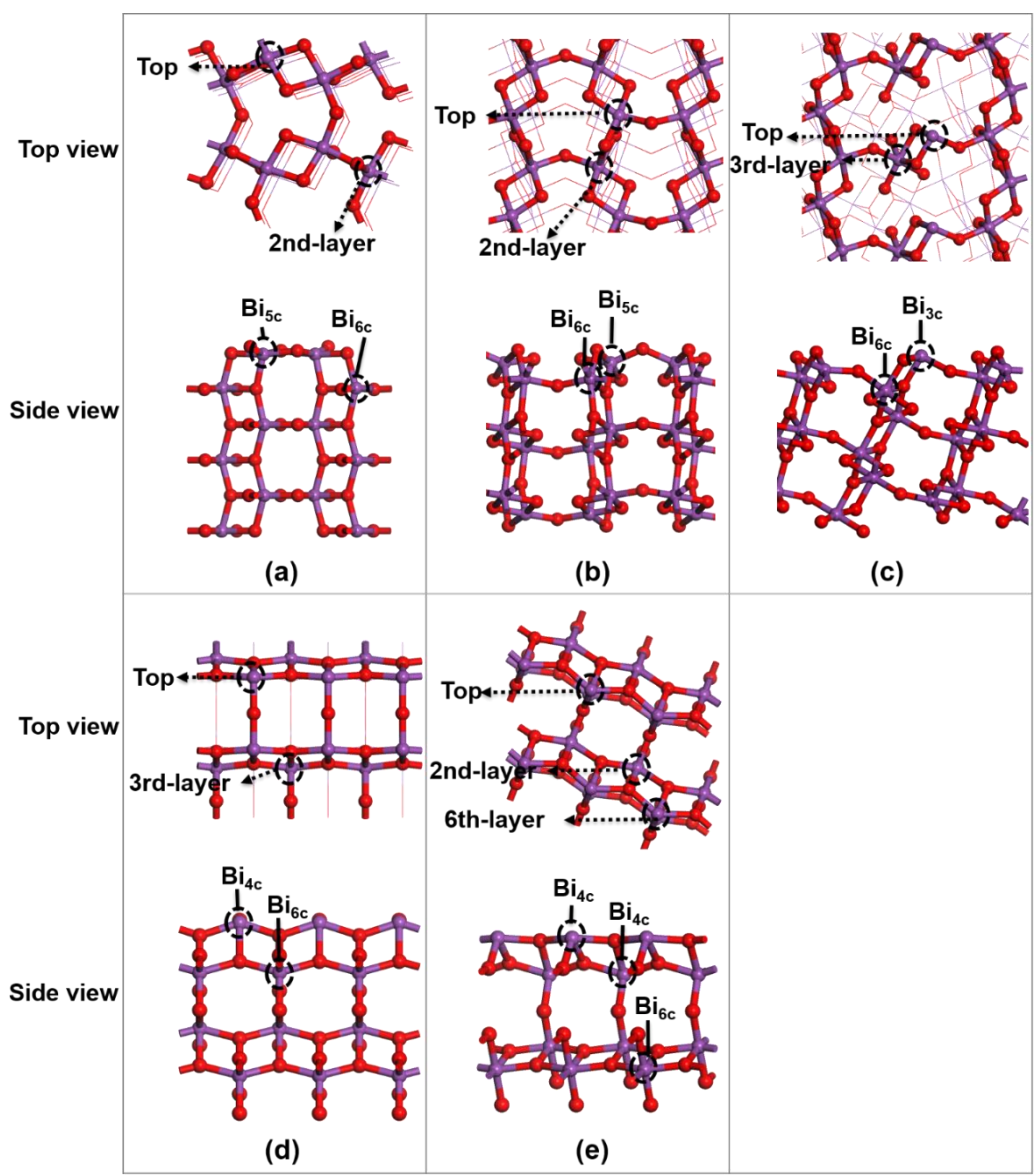

**Figure 3**. Surface morphology and the corresponding adsorption sites of $Bi_2O_5$: (a) (100), (b) (110), (c) (111), (d) (010), and (e) (001) surfaces, respectively. Color legend: Bi purple, O red.

**2.2.4. Additional technical settings**

Within all DFT calculations, a vacuum with 15 Å was set in *c* direction to avoid the interactions between different slab models. Atoms in top two layers were fully relaxed while atoms in the rest layers were fixed. Cornering the adsorption sites, Taylor et al. proposed that the active centers are the positions raised on the surfaces since the atoms at this position are unsaturated, having surface free valence electrons, which have high capacity to adsorb reactant, intermediates, and products. [45] Here, following the theory proposed by Taylor et al., top sites marked in the Figure 1 to 3 were used as the active sites for $CO_2RR$ reaction, i.e. $Bi_{3c}$ in Figure 1 and Figure 2 for all facets of metallic Bi and $Bi_2O_3$, while situation becomes complicated for $Bi_2O_5$, of which $Bi_{3c}$ is the top sites (or reactive sites) for (111), $Bi_{4c}$ for (010) and (001), and $Bi_{5c}$ for (100) and (110).

## 3. RESULTS AND DISCUSSION

### 3.1. $CO_2RR$ over the metallic Bi (0) surface

The stability of catalyst surfaces is an important indicator for their catalytic properties. Usually, two terms are popularly used to define the stability of the surfaces. The first one is the formation energy, of which the higher value means the greater energy required to form such facets, in other words, less stable the surface is (see Eq. 1 for the detail calculations). The second one is the surface energy, of which the lower surface energy is, the more stable facet is, indicating that the crystal tends to grow in such directions, thereby more distributions would be obtained in the experiments (see Eq. 2 for detail calculations). The computed formation and surface energies of (100), (110) and (111) for metallic Bi were given in Figure 4a, and the results show that both surface and formation energies of three low-index facets are relative small, which is consistent with the experimental findings where these three facets are the most accessible ones. [19, 41, 42] Specifically speaking, the formation energies of (100), (110) and (111) are 0.34 eV, 0.50 eV and 0.33 eV, respectively, while surface energies are 0.06 eV, 0.08 eV and 0.05 eV, respectively. As a summary, both formation energies and surface energies reveal that (111) is the most stable facets for the metallic Bi, similar to the report by Hofmann.[46] Afterwards, the 2$e^-$ transferred of $CO_2$RR reactions were investigated on these three low-index crystal faces, with the products being HCOOH and CO, together with the side reaction of HER. According to literatures, the production of HCOOH on Bi-based catalysts is through OCHO* intermediates, while the production of CO is through COOH* intermediates.[20, 47] The computed reaction path is depicted in the Figure b, c and d, and the optimized structures are provided in Figure S1. Generally, our DFT results are well consistent with previous reports.[8, 31, 35, 42, 48] The adsorption energies of $CO_2$ on (100), (110) and (111) surfaces are -0.16eV, -0.17eV and -0.13eV, respectively, indicating week interactions, most contributed by Van der Waals forces, are existing between $CO_2$ and the three low-index crystal faces (Figure 4 b-d). For (100), rate determining steps (RDS) for HCOOH and CO are confirmed to be the elementary step from OCHO* to HCOOH*, and from COOH* to CO*, with the energy barriers being 1.69 eV and 1.45 eV, respectively. The same RDS steps were found for HCOOH and CO in case of (110), with the corresponding energy barriers being 1.92 eV and 1.38 eV, respectively. Different data have been obtained for (111), of which the RDS for CO is the elementary step from $CO_2$ to COOH, while the RDS for HCOOH remains the same to that of (100) and (110). Moreover, the computed energy barriers for both HCOOH and CO are obviously smaller compared to that of (100) and (110), being 0.31 eV and 0.54 eV, respectively. As such, we conclude that (111) is the most reactive facets for metallic Bi. We also computed energy barriers of HER for these three low-index crystal faces, and the results are summarized in Figure S2, and the results show that all energy barriers are extremely high, being 2.49 eV, 3.30 eV and 1.80 eV, for (100), (110) and (111), respectively, indicating that the HER are almost inhibited on all studied metallic Bi surfaces.

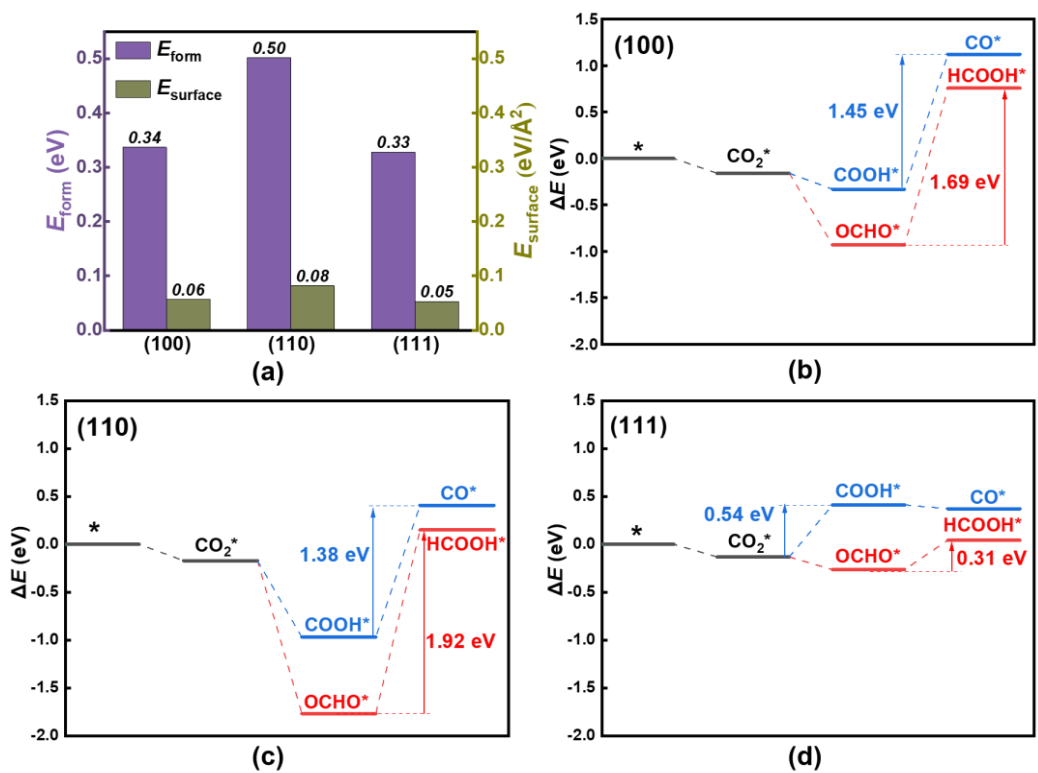

**Figure 4**. (a) The formation and surface energies of the metallic Bi surface facets. The computed reaction path of CO$_2$RR towards HCOOH and CO over (100) (b), (110) (c), and (111) (d) facets.

### 3.2. CO$_2$RR over the Bi (+3) surface

The computed formation and surface energies for (001), (010) and (100) of Bi$_2$O$_3$ are given in Figure 5a. The results show that the formation energies are in general much higher than surface energies. For instant, the computed formation energies are 2.04 eV, 2.42 eV and 2.50 eV, respectively, for (001), (010) and (100), of which the surface energies are only 0.08 eV, 0.11 eV, and 0.08 eV. Moreover, we found that the formation energies of the three low-index crystal facets of Bi$_2$O$_3$ is much higher than that of low-index crystal facets of metallic Bi. These observations might be because of that unlike the metallic Bi, where the layers are connected with only Bi-Bi metallic like bonds, being 3.06 Å, the layers are connected *via* strong Bi-O-Bi covalent bonds in case of bulk Bi$_2$O$_3$ materials. Another reason might be the polarity. Tasker divides the surface of metal oxides into three categories, of which the polar surface with a dipole moment perpendicular to the surface is electrostatically unstable.[49] The chemical bond of surface atoms is usually unsaturated, and surface restructuring is easy to occur in order to achieve saturated coordination. Therefore, the stable existence of the surface of metal oxides is relative difficult than that of metal single crystals. The computed reaction path for all studied facets of $Bi_2O_3$ is give in Figure 5 b to d, and corresponding optimized structures are provided in Figure S3. First, the RDS of HCOOH and CO are all the same, which are the elementary steps from OCHO* to HCCOH, and $CO_2$* to COOH*, respectively. Note that these findings are somehow different from that of the metallic Bi facets, as discussed in the previous section and Figure 4. Moreover, the RDS of HCOOH on (100) and (010) crystal facets are much lower than that for CO (Figure 5b, c), while the RDS for HCOOH on (001) crystal facets is higher than that for CO (Figure 5d), indicating that the selectivity of HCOOH versus CO over (100) and (010) crystal facets are higher that over the (001) crystal facet, in other words, the (100) and (010) crystal facets are most likely to produce HCOOH, while the (001) crystal facet prefers to produce CO. In terms of absolute values of RDS energy barriers, the (010) crystal facet would be best choice for $2e^-$ transfer process, and HCOOH is the main product, since its barrier of RDS for HCOOD is only 0.06 eV. We also examined the HER reactions on all selected surface facets of $Bi_2O_3$, and we found that the HER reaction could be inhibited at all surfaces with an energy barrier being 0.91 eV at least (see Figure S4).

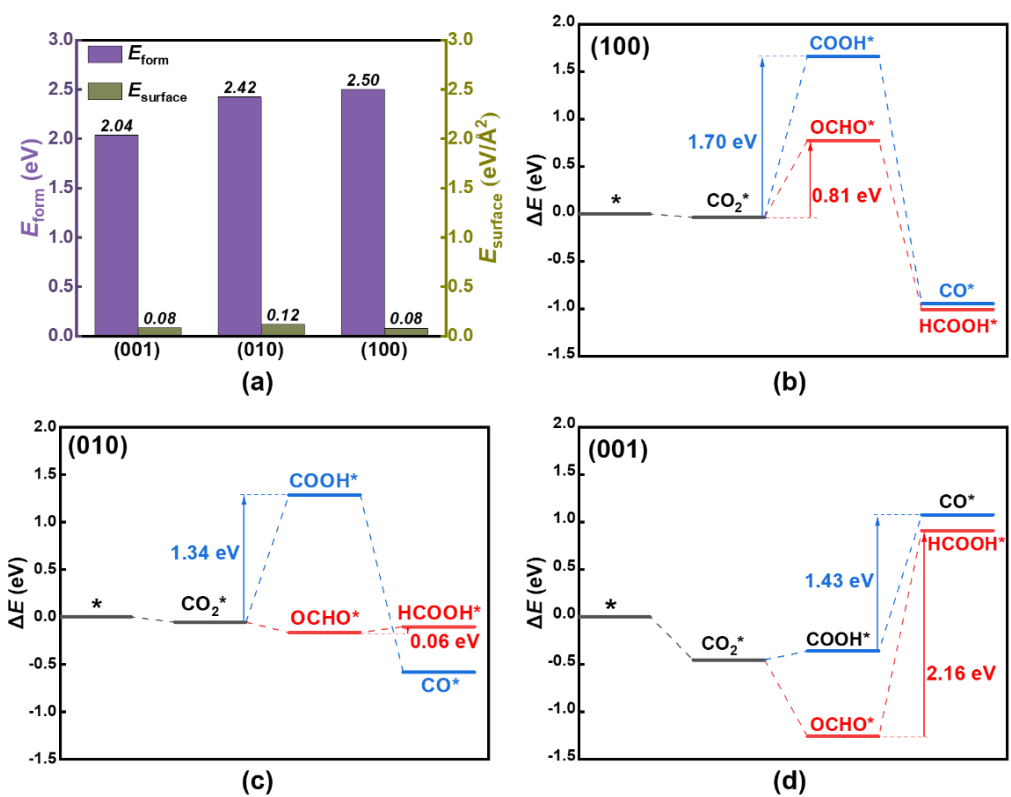

**Figure 5**. (a)The formation and surface energies of the $Bi_2O_3$ surface facets. The computed reaction path of $CO_2RR$ towards HCOOH and CO over (100) (b), (010) (c), and (001) (d) facets.

### 3.3. $CO_2RR$ over the Bi (+5) surface

We then computed formation and surface energies as well as the reaction path for the selected five low-index crystal facets of $Bi_2O_5$, and the results are provided in Figure 6, and the optimized structures are given in Figure S5. Generally, the formation energies for crystal facets of $Bi_2O_5$ are between that of metallic Bi and $Bi_2O_3$, while surface energies of three valence states of Bi are comparable, which is normally ca.0.1 eV. For $Bi_2O_5$ itself, the formation and surface energy of (110) crystal facet are the smallest, which are 0.75 eV and 0.04 eV, respectively, indicating that (110) crystal facet is the most stable and most populated one for $Bi_2O_5$ (Figure 6a). Talking about the reaction pathway, we found that for all studied facets, the RDS for the formation of HCOOH and CO are the same, which are the elementary steps from OCHO* to HCOOH, and from $CO_2$* to COOH* (Figure 6 b-d). However, the energy barriers are relatively high for both HCOOH and CO, with the values of more than 1.00 eV. As such, the selected five low-index crystal faces of $Bi_2O_5$ are not suitable for the 2$e^-$ transfer CO$_2$RR reactions from the theoretical point of view. Moreover, unlike the situation of metallic Bi and Bi$_2$O$_3$, HER is able to occur in the case of Bi$_2$O$_5$. For example, the energy barriers of HER on (100), (110) and (111) crystal facets are much lower than the RDS of HCOOH, and the HER of (010) is only 0.01 eV lower than that of HCOOH (see Figure 6). As such, we conclude that Bi$_2$O$_5$, representing Bi at its +5 valence state, is not an ideal candidate for 2$e^-$ transfer CO$_2$RR reaction, due to the fact that formation of HCOOH and CO have higher energy barriers, and HER reactions could occur with relative low energy barriers.

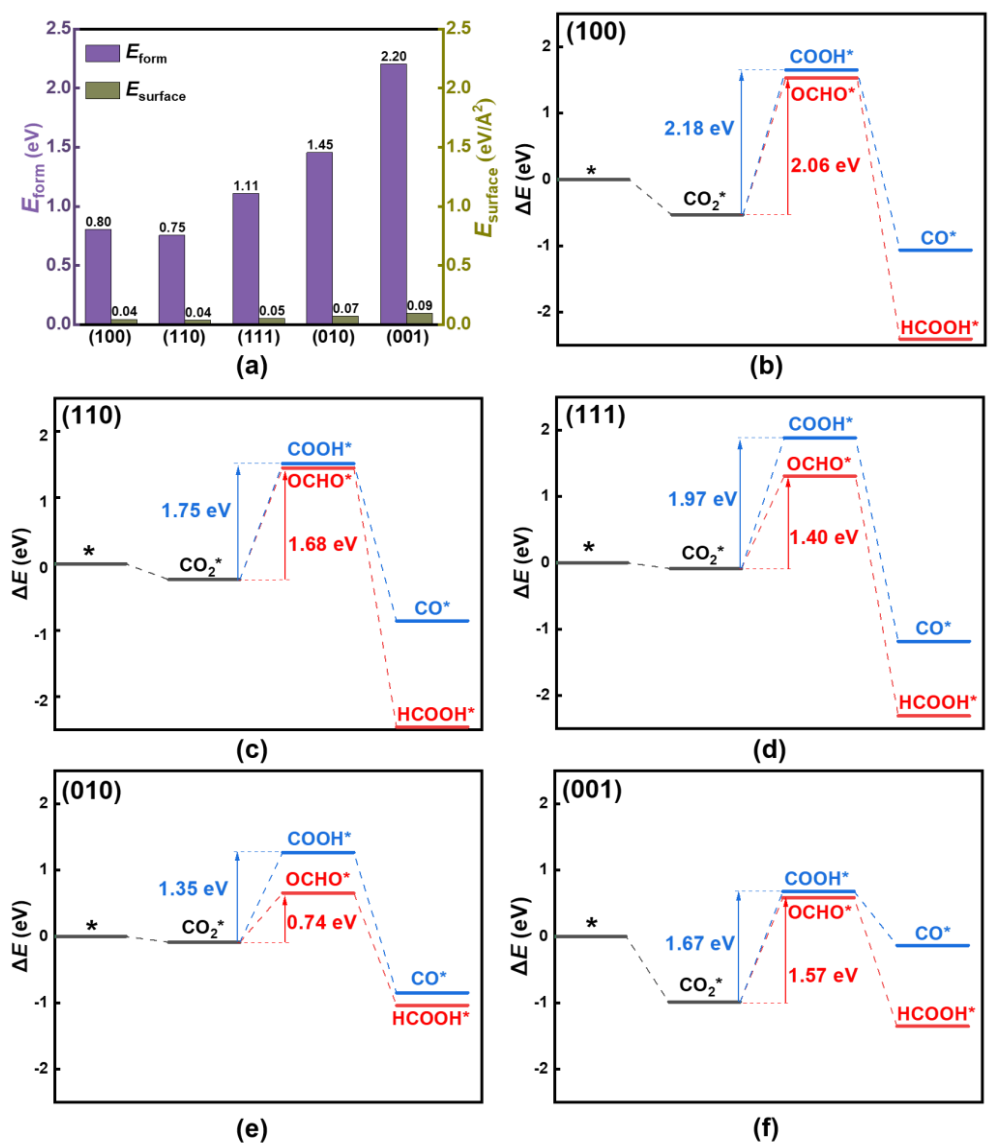

**Figure 6**. (a) The formation and surface energies of Bi$_2$O$_5$ surface facets. The computed reaction path of CO$_2$RR towards HCOOH and CO over (100) (b), (110) (c), (111) (d), (010)

(e), and (001) (f) facets.

As a summary, we found that both metallic Bi and $Bi_2O_3$ are suitable for the reduction of $CO_2$ to HCOOH and CO, while $Bi_2O_5$ is screened out based on our DFT calculations. For metallic Bi, we identify (111) crystal facet is the best choice, of which both energy barriers for the formation of HCOOH and CO are relatively low, being 0.54 eV and 0.31 eV, respectively. For $Bi_2O_3$, we identify (010) crystal facet is the best choice, of which the energy barriers for the formation of HCOOH and CO are 1.34 eV and 0.06 eV, respectively. In terms of absolute values, we conclude that metallic Bi has a higher conversion rate for $CO_2$, while $Bi_2O_3$ has a better selectivity towards the HCOOH.

### 3.4. Post-analysis of the selectivity of HCOOH versus CO

We then compared the adsorption energies of intermediates, OCHO*, which gives the product of HCOOH and COOH*, which gives the product of CO (see Figure 7a). The adsorption energies of OCHO* on Bi (111) and $Bi_2O_3$(010) facets are -0.26 eV and -0.16 eV, respectively, while adsorption energies of COOH* are 0.41 eV and 1.28 eV, respectively. These values are well consistent with the conversion rate of $CO_2$ and selectivity of HCOOH for the Bi (111) and $Bi_2O_3$(010) facets, that is, Bi (111) demonstrates a poor selectivity for HCOOH, but have a higher conversion rate for $CO_2$, while $Bi_2O_3$ (010) exhibits good selectivity for HCOOH production, yet it displays a lower conversion rate for $CO_2$. The analysis of band structures and charge transfer also support our conclusions. Changing from Bi(111) to $Bi_2O_3$, the *d* band center moves to more positive values from -1.62 eV to -1.40 eV (Figure 7b), and less charge transfer have been found between the intermediates and the facets, *i.e.* 0.48 *e* and 0.70 *e* were computed in case of Bi (111) for COOH* and OCHO*, and 0.44 *e* and 0.69 *e* were computed in case of $Bi_2O_3$ (Figure 7c).

To further prove above-mentioned conclusion, we started from Bi (111) facets, and gradually add O into the surface, denoted as 1O-Bi (111), 2O-Bi (111) and 3O-Bi (111), and the computed the reaction path for both HCOOH and CO (see Figure S10 a, b and c for optimized structures and reaction path). Interestingly, we found that with the increase of O content, the conversion rate for $CO_2$ gradually decreases, while the selectivity gradually increases (Figure 7 d). For example, energy barriers for HCOOH gradually decrease from 0.31 eV to 0.06 eV, energy barriers for CO gradually increase from 0.54 eV to 1.34 eV, and energy barriers for HER gradually decrease from 1.80 eV to 0.91 eV, indicating that less $2e^-$ products (HCOOH and CO) would be obtained, in other words, the conversion rates of $CO_2$ decrease. Moreover, the difference between the energy barriers for HCOOH and CO gradually increase from 0.23 eV to 1.28 eV, indicating that selectivity of HCOOH increases.

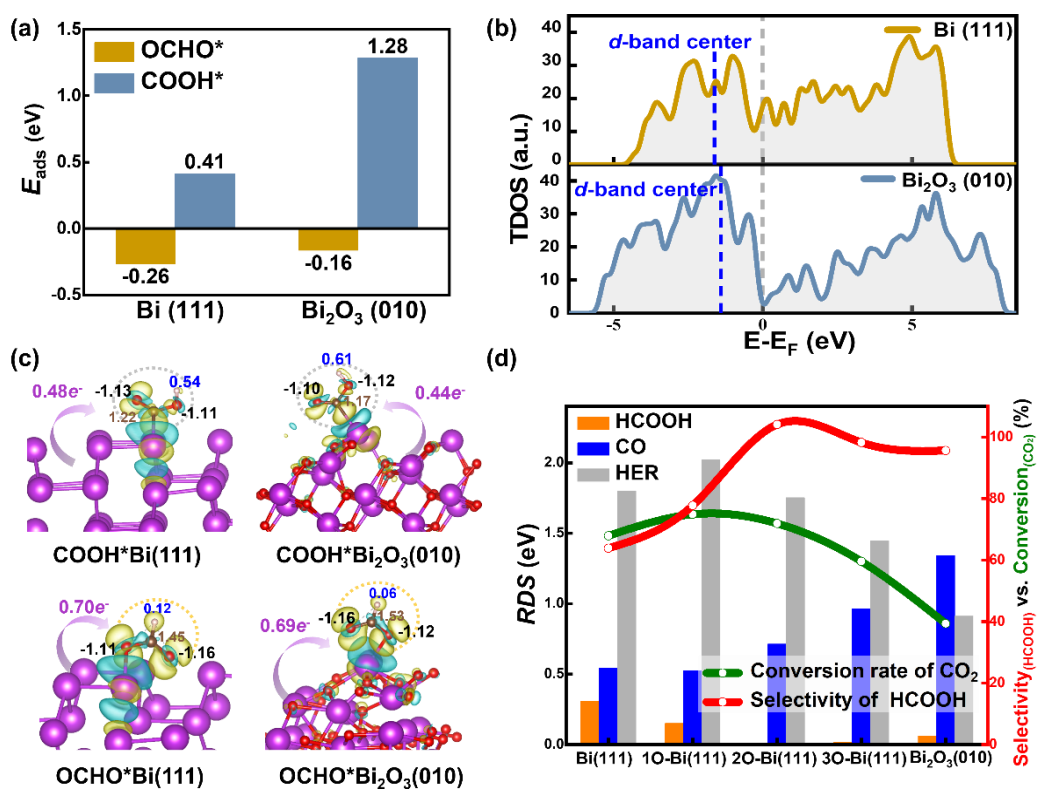

**Figure 7.** (a) The adsorption energies of OCHO* and COOH* on Bi (111) and $Bi_2O_3$(010) crystal facets, (b) The total density of state (TDOS) for Bi (111) and $Bi_2O_3$(010) crystal facets, (c) The charge density difference and Bader charge of the OCHO* and COOH* on Bi (111) and $Bi_2O_3$(010) crystal facets, and (d) The selectivity of HCOOH and conversion of $CO_2$ of bismuth with different oxygen contents.
P
A
G
E

## 5. CONCLUSIONS

Here, we performed a comprehensive theoretical study *via* density functional theory on the effects of valence states and crystal facets on the conversion rate of $CO_2$ and selectivity for the $2e^-$ transfer $CO_2RR$ process, where the HCOOH and CO are the products. Three typical valence states of Bi, such as 0 (Bi), +3 ($Bi_2O_3$) and +5 ($Bi_2O_5$), as well as their often-studied crystal facets were selected for the reaction path calculations. The results show that in terms of individual valence state, the best catalytic surface for metallic Bi is the (111) facets, of which both energy barriers for HCCOH and CO are relatively low, being 0.31 eV and 0.54 eV, respectively. For $Bi_2O_3$, the best catalytic surface is the (010) facets, and the energy barriers for HCCOH and CO are 0.06 eV and 1.34 eV, respectively. For $Bi_2O_5$, that values are 0.74 eV and 1.34 eV, respectively. Taken the HER reactions into account, we screened out the surfaces of $Bi_2O_5$, since the energy barriers of HER are considerably lower than that of the $2e^-$ $CO_2RR$ process, while energy barriers of HER are much higher than that of the $2e^-$ $CO_2RR$ process in all the cases of metallic Bi and $Bi_2O_3$. More importantly, we found that Bi (111) has a poor selectivity for HCOOH, but boasts a higher conversion rate for $CO_2$, which is because both energy barriers are relatively low and almost the same. While $Bi_2O_3$ exhibits a good selectivity for HCOOH production, yet it displays a lower conversion rate for $CO_2$, which is because CO formation has much larger energy barrier compared to that of HCOOH. There observations are then explained by the *d* band center theory and the charge transfer between the intermediates and the surfaces. Subsequently, we further examined the effects of oxygen contents on the selectivity of HCOOH and the conversion rate for $CO_2$. Interestingly, we found with the increase of oxygen contents, the selectivity of HCCOH increases, while the conversion rate of $CO_2$ decease. The best situation is the situation where Bi is partially oxidized. With these observations, we propose that a system which is consisting of the mixture phase of metallic Bi and $Bi_2O_3$ should be considered in future experiments to improve the efficiency for $2e^-$ transfer $CO_2RR$ process.

## ACKNOWLEDGEMENT

This work was supported by National Natural Science Foundation of China (21978294, 22076149), Innovative Team Program of Natural Science Foundation of Hubei Province (2023AFA027), Natural Science Foundation of Hubei Province (No. 244067), and the start-up funding from Wuhan Textile University (No. 20220321).

## CONFLICTS OF INTEREST

The authors declare that they have no known competing financial interests.